# The relation of genuine multiqubit entanglement and controlled teleportation

Xin-Wei Zha, Hai-Yang Song

Department of Applied Mathematics and Applied Physics, Xi'an Institute of Posts and Telecommunications, Xi'an, 710061, China

**Abstract**: Recently, Paolo Facchi et al. [Phys. Rev. A. 77, 060304 (R) (2008)] introduced the notion of maximally multipartite entangled states of n qubits. Here, we give a criterion for faithful controlled teleportation of an arbitrary two-qubit state via a five-qubit entangled state and obtain the general relation between the genuine five-qubit entangled state and controlled teleportation. This criterion can be extended to teleportation of an arbitrary N-qubit state using 2N+1-qubit entangled state. Furthermore, we study the optimal match of measuring basis and quantum channel.

Quantum teleportation is a prime example of a quantum information processing task, where an unknown state can be perfectly transported from one place to another using previously shared entanglement and classical communication between the sender and the receiver. Since the first creation of quantum teleportation protocol by Bennett [1], research on quantum teleportation has been attracting much attention both theoretically and experimentally in recent years due to its important applications in quantum calculation and quantum communication. A number of remarkable theoretical concepts and schemes have also been invented for multi-particle teleportation and controlled teleportation [2-11]. Recently, Man et al [12] present an explicit genuine 2N+1-qubit entangled state motivated by the so-called controlled teleportation of an unknown N-qubit state. In their paper, they gave the expression of genuine five-qubit entangled state, which is defined as

$$|M_5\rangle_{12345} = \frac{1}{4}(|00000\rangle + |00001\rangle + |00110\rangle - |00111\rangle$$
$$+ |01010\rangle + |01011\rangle - |01100\rangle + |01101\rangle$$
$$- |10010\rangle + |10011\rangle + |10100\rangle + |10101\rangle$$
$$+ |11000\rangle - |11001\rangle + |11110\rangle + |11111\rangle)_{12345} \quad (1)$$

For their state (1), calculation yield $Tr(\rho_{j_1 j_2}^2) = \frac{1}{4}$, for $j_1 j_2 \in \{12, 23, 34, 45, 35, 14, 25, 15\}$, $Tr(\rho_{k_1 k_2}^2) = \frac{1}{2}$ for $k_1 k_2 \in \{13, 24\}$. However, Paolo Facchi et al [13] introduce the notion of maximally multipartite entangled states of n qubits as a generalization of the bipartite case. For five-qubit $\pi_{ME} = \frac{1}{4}$ i.e. $Tr(\rho_{ij}^2) = \frac{1}{4}$ for $ij \in \{12,13,14,15,23,24,25,34,35,45\}$. Therefore, According to definition of Ref.[13], Man et al. gave the expression of five-qubit entangled state is not maximally multipartite entangled states. In this paper, we present a criterion for faithful controlled teleportation of an arbitrary two-qubit state via a five-qubit entangled state and obtain the general relation between the genuine five-qubit entangled state and controlled teleportation.

In order to obtain the criterion for faithful controlled teleportation we suppose that the sender Alice has two particles $a_1, a_2$ in an unknown state:

$$|\chi\rangle_{a_1 a_2} = (x_0|00\rangle + x_1|01\rangle + x_2|10\rangle + x_3|11\rangle)_{a_1 a_2} \tag{2}$$

where $x_0$, $x_1$, $x_2$ and $x_3$ are arbitrary complex numbers, and it is assumed that the wave function satisfies the normalization condition $\sum_{i=0}^{3}|x_i|^2 = 1$. Alice, Bob, and Charlie share beforehand a quantum channel of the form

$$\begin{aligned}|M_5\rangle_{A_1 A_2 B_1 B_2 C} = &(a_0|00000\rangle + a_1|00001\rangle + a_2|00010\rangle + a_3|00011\rangle \\ &+ a_4|00100\rangle + a_5|00101\rangle + a_6|00110\rangle + a_7|00111\rangle \\ &+ a_8|01000\rangle + a_9|01001\rangle + a_{10}|01010\rangle + a_{11}|01011\rangle \\ &+ a_{12}|01100\rangle + a_{13}|01101\rangle + a_{14}|01110\rangle + a_{15}|01111\rangle \\ &+ a_{16}|10000\rangle + a_{17}|10001\rangle + a_{18}|10010\rangle + a_{19}|10011\rangle \\ &+ a_{20}|10100\rangle + a_{21}|10101\rangle + a_{22}|10110\rangle + a_{23}|10111\rangle \\ &+ a_{24}|11000\rangle + a_{25}|11001\rangle + a_{26}|11010\rangle + a_{27}|11011\rangle \\ &+ a_{28}|11100\rangle + a_{29}|11101\rangle + a_{30}|11110\rangle + a_{31}|01111\rangle)_{A_1 A_2 B_1 B_2 C}\end{aligned} \tag{3}$$

Therefore, the joint state of the whole system can be expressed as:

$$|\psi_s\rangle = |\chi\rangle_{a_1 a_2}|M_5\rangle_{A_1 A_2 B_1 B_2 C} \tag{4a}$$

In order to realize teleportation, firstly, Alice has to perform Bell-state measurements on qubit pairs $(a_1, A_1), (a_2, A_2)$, Subsequently, Charlie performs a Von Neumann measurement on his single qubit then Bob can perform a corresponding unitary transformation to reconstruct original state in particle $(B_1, B_2)$.

In accordance with the principle of superposition, $|\psi\rangle_s$ can be represented in the following form [15].

$$|\psi_s\rangle = |\chi\rangle_{a_1 a_2} |\psi\rangle_{A_1 A_2 B_1 B_2 C} = \frac{1}{4\sqrt{2}} \sum_{i=1}^{4} \sum_{j=1}^{4} \sum_{n=1}^{2} |\varphi^i_{a_1 A_1}\rangle |\varphi^j_{a_2 A_2}\rangle |\varphi^n_C\rangle \hat{\sigma}^{ijn}_{B_1 B_2} |\chi\rangle_{B_1 B_2} \qquad (4b)$$

where $|\varphi^i_{a_1 A_1}\rangle, |\varphi^j_{a_2 A_2}\rangle$ are Bell states, and

$$|\varphi^1_C\rangle = \cos\theta |0\rangle + \sin\theta |1\rangle$$

$$|\varphi^2_C\rangle = \sin\theta |0\rangle - \cos\theta |1\rangle$$

$$|\chi\rangle_{B_1 B_2} = (x_0 |00\rangle + x_1 |01\rangle + x_2 |10\rangle + x_3 |11\rangle)_{B_1 B_2}.$$

The operator $\hat{\sigma}^{ijn}_{B_1 B_2}$ here is called the "transformation operator". The criterion for faithfully teleporting an arbitrary two-qubit state can be given in terms of the "transformation operator". If $\hat{\sigma}^{ijn}_{B_1 B_2}$ is a unitary operator, Bob can determine the state of particles $(B_1, B_2)$ exactly by the inverse of the transformation operator $(\hat{\sigma}^{ijn}_{B_1 B_2})^{-1}$, and

$$\hat{\sigma}^{ij1}_{B_1 B_2} = \hat{\sigma}^{111}_{B_1 B_2} \left( \sigma^i_{B_1} \otimes \sigma^j_{B_2} \right), \qquad \hat{\sigma}^{ij2}_{B_1 B_2} = \hat{\sigma}^{112}_{B_1 B_2} \left( \sigma^i_{B_1} \otimes \sigma^j_{B_2} \right), \quad i, j = 1, 2, 3, 4 \qquad (5)$$

where $\hat{\sigma}^k_m = I_m, \sigma_{mz}, \sigma_{mx}, -i\sigma_{my}$ $m = B_1, B_2$, $I_m$ is the two-dimensional identity and $\sigma_{mz}, \sigma_{mx}, \sigma_{my}$ are the Pauli matrices. Apparently, if $\hat{\sigma}^{111}_{B_1 B_2}, \hat{\sigma}^{112}_{B_1 B_2}$ is a unitary operator, $\hat{\sigma}^{ijn}_{B_1 B_2}$ are also unitary operators.

From Eq. (4), we can easily obtain transformation operator

$$\hat{\sigma}^{111}_{B_1 B_2} = \langle \varphi^1_C | \langle \varphi^1_{a_1 A_1} | \langle \varphi^1_{a_2 A_2} | | \psi_s\rangle$$

$$= 2\sqrt{2} \begin{pmatrix} a_0 \cos\theta + a_1 \sin\theta & a_2 \cos\theta + a_3 \sin\theta & a_4 \cos\theta + a_5 \sin\theta & a_6 \cos\theta + a_7 \sin\theta \\ a_8 \cos\theta + a_9 \sin\theta & a_{10} \cos\theta + a_{11} \sin\theta & a_{12} \cos\theta + a_{13} \sin\theta & a_{14} \cos\theta + a_{15} \sin\theta \\ a_{16} \cos\theta + a_{17} \sin\theta & a_{18} \cos\theta + a_{19} \sin\theta & a_{20} \cos\theta + a_{21} \sin\theta & a_{22} \cos\theta + a_{23} \sin\theta \\ a_{24} \cos\theta + a_{25} \sin\theta & a_{26} \cos\theta + a_{27} \sin\theta & a_{28} \cos\theta + a_{29} \sin\theta & a_{30} \cos\theta + a_{31} \sin\theta \end{pmatrix}$$

(6a)

$$\hat{\sigma}^{112}_{B_1 B_2} = \langle \varphi^2_C | \langle \varphi^1_{a_1 A_1} | \langle \varphi^1_{a_2 A_2} | | \psi_s\rangle$$

$$= 2\sqrt{2} \begin{pmatrix} a_0 \sin\theta - a_1 \cos\theta & a_2 \sin\theta - a_3 \cos\theta & a_4 \sin\theta - a_5 \cos\theta & a_6 \sin\theta - a_7 \cos\theta \\ a_8 \sin\theta - a_9 \cos\theta & a_{10} \sin\theta - a_{11} \cos\theta & a_{12} \sin\theta - a_{13} \cos\theta & a_{14} \sin\theta - a_{15} \cos\theta \\ a_{16} \sin\theta - a_{17} \cos\theta & a_{18} \sin\theta - a_{19} \cos\theta & a_{20} \sin\theta - a_{21} \cos\theta & a_{22} \sin\theta - a_{23} \cos\theta \\ a_{24} \sin\theta - a_{25} \cos\theta & a_{26} \sin\theta - a_{27} \cos\theta & a_{28} \sin\theta - a_{29} \cos\theta & a_{30} \sin\theta - a_{31} \cos\theta \end{pmatrix}$$

(6b)

Now let us assume, $\hat{\sigma}_{B_1B_2}^{111}$, $\hat{\sigma}_{B_1B_2}^{112}$ is a unitary operator, from Eqs. (6a) and (6b), we can obtain:

$$a_0 a_8^* + a_1 a_9^* + a_2 a_{10}^* + a_3 a_{11}^* + a_4 a_{12}^* + a_5 a_{13}^* + a_6 a_{14}^* + a_7 a_{15}^* = 0, \cdots,$$

$$|a_0|^2 + |a_1|^2 + |a_2|^2 + |a_3|^2 + |a_4|^2 + |a_5|^2 + |a_6|^2 + |a_7|^2 = \frac{1}{4}, \cdots,, \quad (7)$$

On the other hand, we have the entanglement measure [14]

$$\pi_{12} = Tr_{12} \rho_{12}^2$$
$$= \left(|a_0|^2 + |a_1|^2 + |a_2|^2 + |a_3|^2 + |a_4|^2 + |a_5|^2 + |a_6|^2 + |a_7|^2\right)^2$$
$$+ \left(|a_8|^2 + |a_9|^2 + |a_{10}|^2 + |a_{11}|^2 + |a_{12}|^2 + |a_{13}|^2 + |a_{14}|^2 + |a_{15}|^2\right)^2$$
$$+ \left(|a_{16}|^2 + |a_{17}|^2 + |a_{18}|^2 + |a_{19}|^2 + |a_{20}|^2 + |a_{21}|^2 + |a_{22}|^2 + |a_{23}|^2\right)^2$$
$$+ \left(|a_{24}|^2 + |a_{25}|^2 + |a_{26}|^2 + |a_{27}|^2 + |a_{28}|^2 + |a_{29}|^2 + |a_{30}|^2 + |a_{31}|^2\right)^2$$
$$+ 2\left|a_0 a_8^* + a_1 a_9^* + a_2 a_{10}^* + a_3 a_{11}^* + a_4 a_{12}^* + a_5 a_{13}^* + a_6 a_{14}^* + a_7 a_{15}^*\right|^2$$
$$+ 2\left|a_0 a_{16}^* + a_1 a_{17}^* + a_2 a_{18}^* + a_3 a_{19}^* + a_4 a_{20}^* + a_5 a_{21}^* + a_6 a_{22}^* + a_7 a_{23}^*\right|^2$$
$$+ 2\left|a_0 a_{24}^* + a_1 a_{25}^* + a_2 a_{26}^* + a_3 a_{27}^* + a_4 a_{28}^* + a_5 a_{29}^* + a_6 a_{30}^* + a_7 a_{31}^*\right|^2$$
$$+ 2\left|a_8 a_{16}^* + a_9 a_{17}^* + a_{10} a_{18}^* + a_{11} a_{19}^* + a_{12} a_{20}^* + a_{13} a_{21}^* + a_{14} a_{22}^* + a_{15} a_{23}^*\right|^2$$
$$+ 2\left|a_8 a_{24}^* + a_9 a_{25}^* + a_{10} a_{26}^* + a_{11} a_{27}^* + a_{12} a_{28}^* + a_{13} a_{29}^* + a_{14} a_{30}^* + a_{15} a_{31}^*\right|^2$$
$$+ 2\left|a_{16} a_{24}^* + a_{17} a_{25}^* + a_{18} a_{26}^* + a_{19} a_{27}^* + a_{20} a_{28}^* + a_{21} a_{29}^* + a_{22} a_{30}^* + a_{23} a_{31}^*\right|^2 \quad (8)$$

Since $\hat{\sigma}_{B_1B_2}^{111}$, $\hat{\sigma}_{B_1B_2}^{112}$ is a unitary operator, thus the entanglement measure can be expressed

$\pi_{12} = \pi_{A_1A_2} = \frac{1}{4}$. Analogous, we can obtain $\pi_{34} = \pi_{B_1B_2} = \frac{1}{4}$. Therefore, if the state as a quantum channel can realize faithful controlled teleportation of an arbitrary two-qubit state, then it must have $\pi_{12} = \pi_{A_1A_2} = \frac{1}{4}, \pi_{34} = \pi_{B_1B_2} = \frac{1}{4}$. It should be noted that the maximally entanglement five-qubit have $Tr(\rho_{ij}^2) = \frac{1}{4}$ for $ij \in \{12,13,14,15,23,24,25,34,35,45\}$. For maximally multipartite entangled five-qubit state, no matter how the qubits are partitioned between Alice, Bob and Charlie, the unknown two-particle entangled state controlled teleportation can be realized perfectly, and the successful possibilities and the fidelities both reach unity.

As an example, we now consider a maximally five-qubit entanglement state (Brown state [15]), this state has the form

$$|\psi_5\rangle_{12345} = \frac{1}{2}(|001\rangle|\varphi_-\rangle + |010\rangle|\psi_-\rangle + |100\rangle|\varphi_+\rangle + |111\rangle|\psi_+\rangle)_{12345} \qquad (9)$$

where,

$$|\psi_\pm\rangle = \frac{1}{\sqrt{2}}(|00\rangle \pm |11\rangle),$$

$$|\varphi_\pm\rangle = \frac{1}{\sqrt{2}}(|01\rangle \pm |10\rangle), \qquad (10)$$

Substituting formulas (10) into Eq.(9), Eq. (9) can be rewritten as

$$|\psi_5\rangle_{12345} = \frac{1}{2\sqrt{2}}(|00101\rangle - |00110\rangle + |01000\rangle - |01011\rangle \\ + |10001\rangle + |10010\rangle + |11100\rangle + |11111\rangle)_{12345} \qquad (11)$$

If Alice, Bob, and Charlie have particles $12, 34, 5$, the quantum channe can be expressed as:

$$|\psi_5\rangle_{A_1A_2B_1B_2C} = \frac{1}{2\sqrt{2}}(|00101\rangle - |00110\rangle + |01000\rangle - |01011\rangle \\ + |10001\rangle + |10010\rangle + |11100\rangle + |11111\rangle)_{A_1A_2B_1B_2C} \qquad (12a)$$

Using Eqs. (12a), we can express Eq. (6a) and (6b) as

$$\hat{\sigma}^{111}_{B_1B_2} = \begin{pmatrix} 0 & 0 & \sin\theta & -\cos\theta \\ \cos\theta & -\sin\theta & 0 & 0 \\ \sin\theta & \cos\theta & 0 & 0 \\ 0 & 0 & \cos\theta & \sin\theta \end{pmatrix}, \quad \hat{\sigma}^{112}_{B_1B_2} = \begin{pmatrix} 0 & 0 & -\cos\theta & -\sin\theta \\ \sin\theta & \cos\theta & 0 & 0 \\ -\cos\theta & \sin\theta & 0 & 0 \\ 0 & 0 & \sin\theta & -\cos\theta \end{pmatrix}$$

Note that the $\hat{\sigma}^{111}_{B_1B_2}$, $\hat{\sigma}^{112}_{B_1B_2}$ is a unitary operator for any $\theta$.

If Alice, Bob, and Charlie have particle $13, 24, 5$, the quantum channe can be expressed as:

$$|\psi_5\rangle_{A_1A_2B_1B_2C} = \frac{1}{2\sqrt{2}}(|01001\rangle - |01010\rangle + |00100\rangle - |00111\rangle \\ + |10001\rangle + |10010\rangle + |11100\rangle + |11111\rangle)_{A_1A_2B_1B_2C} \qquad (12b)$$

By using Eqs.（6a）and（6b）, we have

$$\hat{\sigma}^{111}_{B_1B_2} = \begin{pmatrix} 0 & 0 & \cos\theta & -\sin\theta \\ \sin\theta & -\cos\theta & 0 & 0 \\ \sin\theta & \cos\theta & 0 & 0 \\ 0 & 0 & \cos\theta & \sin\theta \end{pmatrix}$$

$$\hat{\sigma}_{B_1B_2}^{112} = \begin{pmatrix} 0 & 0 & \sin\theta & \cos\theta \\ -\cos\theta & -\sin\theta & 0 & 0 \\ -\cos\theta & \sin\theta & 0 & 0 \\ 0 & 0 & \sin\theta & -\cos\theta \end{pmatrix}$$

Note that the $\hat{\sigma}_{B_1B_2}^{111}$, $\hat{\sigma}_{B_1B_2}^{112}$ is a unitary operator for any $\theta$.

If Alice, Bob, and Charlie have particle $14, 23, 5$,

$$|\psi_5\rangle_{A_1A_2B_1B_2C} = \frac{1}{2\sqrt{2}}(|00011\rangle - |01010\rangle + |00100\rangle - |01101\rangle \\ + |10001\rangle + |11000\rangle + |10110\rangle + |11111\rangle)_{A_1A_2B_1B_2C} \quad (12c)$$

By using Eqs. (14a) and (14b),

$$\hat{\sigma}_{B_1B_2}^{111} = \begin{pmatrix} 0 & \sin\theta & \cos\theta & 0 \\ 0 & -\cos\theta & -\sin\theta & 0 \\ \sin\theta & 0 & 0 & \cos\theta \\ \cos\theta & 0 & 0 & \sin\theta \end{pmatrix}$$

$$\hat{\sigma}_{B_1B_2}^{112} = \begin{pmatrix} 0 & -\cos\theta & \sin\theta & 0 \\ 0 & -\sin\theta & \cos\theta & 0 \\ -\cos\theta & 0 & 0 & \sin\theta \\ \sin\theta & 0 & 0 & -\cos\theta \end{pmatrix}$$

Obviously, if the $\hat{\sigma}_{B_1B_2}^{111}$, $\hat{\sigma}_{B_1B_2}^{112}$ is a unitary operator, then they must satisfy $\sin\theta\cos\theta = 0$.

In summary, we present a criterion for faithful controlled teleportation of an arbitrary two-qubit state by five qubit-entangled state. According to the criterion, we found that the expression of five-qubit entangled state presented by Man et al [12] is not maximally multipartite entangled states. This criterion can be extended to teleportation of an arbitrary N-qubit state using 2N+1-qubit entangled state. Furthermore, the relation of genuine five-qubit entangled states and faithful controlled teleportation are given. At same time, by transformation operator $\hat{\sigma}_{B_1B_2}^{111}$, $\hat{\sigma}_{B_1B_2}^{112}$, the optimal match of measuring basis and quantum channel is also to be studied.

## Acknowledgements

This work is supported by the National Natural Science Foundation of China (Grant No. 10902083) and the Natural Science Foundation of Shannxi Province (Grant No. 2009GM1007).